\documentclass[10pt, prd, twocolumn, showpacs]{revtex4}

\usepackage{amsmath}
\usepackage{amssymb}
\usepackage{graphicx}
\usepackage{bm} 

\newcommand{\g}[0]{\gamma} 
\newcommand{\al}[0]{\alpha} 
\newcommand{\be}[0]{\beta}
\newcommand{\de}[0]{\delta}
 
\newcommand{\dd}[0]{\partial}

\begin{document}

\title{Thermodynamics, entropy,
and stability of thin shells in 2+1 flat spacetimes}
\author{Jos\'{e} P. S. Lemos}
\email{joselemos@ist.utl.pt}
\affiliation{Centro Multidisciplinar de Astrof\'{\i}sica, CENTRA,
Departamento de F\'{\i}sica, Instituto Superior T\'ecnico - IST,
Universidade de Lisboa - UL, Avenida Rovisco Pais 1,
1049-001 Lisboa, Portugal\,\,}
\author{Gon\c{c}alo M. Quinta}
\email{goncalo.quinta@ist.utl.pt}
\affiliation{Centro Multidisciplinar de Astrof\'{\i}sica, CENTRA,
Departamento de F\'{\i}sica, Instituto Superior T\'ecnico - IST,
Universidade de Lisboa - UL, Avenida Rovisco Pais 1,
1049-001 Lisboa, Portugal\,\,}

\begin{abstract}
The thermodynamic equilibrium states of a static thin ring shell in a
(2+1)-dimensional flat spacetime is analyzed. Inside the ring the
spacetime is flat, whereas outside it is conical flat. The first law
of thermodynamics applied to the thin shell leads to a shell's entropy
which is a function of its mass alone. Two simple forms for this mass
function are given leading to two different expressions for the
entropy. The equations of thermodynamic stability are analyzed
resulting in certain allowed regions for the free parameters. Contrary
to the usual (3+1)-dimensional case this shell's entropy is purely
classic, as the only fundamental constant that enters into the
problem is the (2+1)-dimensional gravitational constant $G_3$, which
has units of inverse mass.

\end{abstract}

\keywords{quasi-black holes, black holes, wormholes one two three}
\pacs{04.40.-b, 04.70.Dy}
\maketitle



\newpage

\section{Introduction}

The study of thermodynamics of gravitational systems has proved
important in many respects, such as the understanding that these
systems can have negative specific heat, and that the gravitational
object par excellence, the black hole, has definite temperature and
entropy associated with it (see, e.g., \cite{york1}), which in turn
indicates that a quantum theory of gravity might be in sight.  In
particular, a special kind of gravitating system, a thin shell
with its surrounding spacetime, is
prone to a direct attack of its thermodynamic properties. Indeed,
Davies, Ford and Page \cite{daviesfordpage} and Hiscock \cite{hiscock},
have shown the usefulness of studying thin shells in (3+1)-dimensional
general relativistic spacetimes from a thermodynamic viewpoint.
Further progress was achieved by Martinez \cite{Mart}, in which
several thermodynamic quantities 
are discussed and a stability analysis of thin
shells is performed using the formalism set in \cite{callen}. For
related studies of thermodynamics of gravitating 
matter, especially on the verge of
becoming a black hole, see \cite{lemoszaslavski1,lemoszaslavskii2}.

Although (3+1)-dimensional spacetimes are of the greatest interest, it
is also pertinent to study thin shells and their properties in
(2+1)-dimensional spacetimes.  The relevance in the study of 
three-dimensional gravity
started with the work of Deser, Jackiw, and 't Hooft \cite{djt}, where
it was shown that, though the corresponding vacuum solution is trivial
since it consists of Minkowski spacetime, a point particle distorts 
it
into a conical space with the particle 
being located at the vertex of the cone, and
moreover, 
moving point particles display nontrivial dynamics. The next simplest
object, beyond a point a particle in 2+1 dimensions, is a thin shell,
i.e, a ring dividing two vacua regions, the interior and exterior to
the ring itself. A detailed analysis of the dynamics of such a ring, 
collapsing or expanding was performed in \cite{mannoh}, 
where the usual junction condition formalism \cite{Israel}
was used.

To further understand the thermodynamic properties of
gravitational systems we propose to study the thermodynamics of a
static thin ring shell in (2+1)-dimensional  general relativity.  By
making use of the appropriate junction conditions for general
relativity \cite{Israel}, one can determine the pressure and rest mass
of the shell in order for it to be static with interior and exterior
spacetimes both flat.  Using then the formalism developed by Martinez
\cite{Mart}, with the thermodynamic theory as presented in
\cite{callen}, one can find generic expressions for the shell's entropy,
which upon some minimal assumptions about the structure of the matter
fields making the system, i.e., an ansatz for the shell temperature in
terms of the gravitational quantities that characterize the system,
yields a definite expression for the entropy of the shell,
and permits a stability analysis.

The paper is organized as follows.  In Sec.~\ref{thinsh}, we will
compute the components of the extrinsic curvature of the shell that
will lead to the shell's linear density and pressure. 
In Sec.~\ref{thermo} we will use these results by directly
inserting them in the first law of thermodynamics to obtain the
differential of the entropy, which will naturally have a degree of
freedom parametrized by an arbitrary function associated with the matter
fields that make up the shell. Two phenomenological expressions will
be considered for the arbitrary function, namely a simple power law of
the rest mass and a power law of a quadratic expression of the rest
mass, which will lead to two different expressions for the
entropy. In Sec.~\ref{eqstate} we will analyze the
thermodynamic stability of the system for each entropy 
function  by
calculating the allowed intervals of the free parameters in order for
the shell to remain thermodynamically stable.
In Sec.~\ref{conc} we  conclude.

\section{The thin shell spacetime}
\label{thinsh}

In 2+1 dimensions Einstein's equation takes the form
\begin{equation}\label{efe} 
G_{ab} = 8\pi G_3 T_{ab}\,, 
\end{equation}
where $G_{ab}$ is the (2+1)-dimensional Einstein tensor, $T_{ab}$ is
the 
stress-energy tensor, and $G_3$ is the gravitational constant in 2+1
dimensions.  $G_3$ has units of inverse mass.  The speed of light is
taken to be equal to $1$.

To find the solution for a thin shell in a (2+1)-dimensional
spacetime, we follow \cite{Israel} and start by considering a 
one-dimensional timelike hypersurface $\Sigma$ 
that partitions spacetime
into two spherically symmetric regions: an inner region
$\mathcal{V}_-$ and an outer region $\mathcal{V}_+$.

Inside the hypersurface we will use flat-polar coordinates
$(t,r,\theta)$ for a flat metric with line element
\begin{equation}\label{LEI}
ds^2 = -dt^2 + dr^2 + r^2 d\theta^2\,, \quad r<R\,,
\end{equation}
where $r=R$ is the radius of the thin-shell hypersurface.

On the outside of the shell, the spacetime is again flat but the
presence of matter justifies the use of conical-polar coordinates
$(t,r,\theta)$, allowing the line element to be written as $ds^2 =
-\beta^2 dt^2 + dr^2 + r^2 \al^2 d\,\theta^2$, for $r>R$, and some
constant $\al$.  It is preferable to make the change $\al\,r\to r$,
and without loss of generality one can put $\beta=\alpha$, so that the
metric takes the form
\begin{equation}
\label{LEO}
ds^2 = - \alpha^2 
dt^2 + \frac{dr^2}{\al^2} + r^2  
d\,\theta^2\,, \quad r>R\,.
\end{equation}
The metric (\ref{LEO}) has a conical singularity at $r=0$ if the
thin-shell hypersurface has a radius $R(\tau) \to 0$, i.e., it turns
into a point particle.

The parametric equations of the thin shell hypersurface $\Sigma$ are
described by $r=R(\tau)$, $t = T(\tau)$, where $\tau$ is the proper
time on the thin shell hypersurface.  Choosing coordinates
$(\tau,\theta)$, we have an induced metric $h_{ab}$ given by
\begin{equation}
ds_{\Sigma}^2 = -d\tau^2 + R^2(\tau) d\theta^2.
\end{equation}

The first junction condition states that in order for $\mathcal{V}_+$
and $\mathcal{V}_-$ to be joined smoothly at $\Sigma$, the induced
metric seen from both sides must satisfy $[h_{ab}]=0$, where the
parentheses symbolize the jump in the metric across the
hypersurface. This condition leads to the relations
\begin{equation} \label{J1}
\al^2 \dot{T}^2 - \frac{\dot{R}^2}{\al^2} = \dot{T}^2 - \dot{R}^2 = 1\,
\end{equation}
where the dot denotes differentiation with respect to $\tau$. 

The
second junction condition involves the extrinsic curvature
$K^{a}{}_{b}$ defined as
\begin{equation}
K^{a}{}_{b} = h^{ac}h^{d}{}_{b}\nabla_{c} n_{d}\,
\end{equation}
where $\nabla_{c}$ denotes covariant derivation.
In general, when there is
a thin matter shell,
the boundary stress-energy tensor $S^{a}{}_{b}$
is related to the jump in extrinsic
curvature by
\begin{equation}
S^{a}{}_{b}=-\frac{1}{8 \pi G_3}\left([K^{a}{}_{b}]-[K]h^{a}{}_{b}\right)
\end{equation}
where $K = h^{b}{}_{a} K^{a}{}_{b}$ and $G_3$ is the gravitational
constant in 2+1 dimensions. From the line elements 
given in Eqs.~(\ref{LEI})
and (\ref{LEO}) and using Eq.~(\ref{J1}), one can compute the
nonzero components of $K^{a}{}_{b}$,
\begin{align}
K^{\tau}_{+}{}_{\tau} &= \frac{\ddot{R}}{\sqrt{\al^2+\dot{R}^2}}\,, \\
K^{\tau}_{-}{}_{\tau} &= \frac{\ddot{R}}{\sqrt{1+\dot{R}^2}}\,, \\
K^{\theta}_{+}{}_{\theta} &= \frac{1}{R}\sqrt{\al^2+\dot{R}^2}\,, \\
K^{\theta}_{-}{}_{\theta} &= \frac{1}{R}\sqrt{1+\dot{R}^2}\,,
\end{align}
which, in turn,
lead to the nonzero components of the linear
boundary stress-energy tensor
\begin{align}
S^{\tau}{}_{\tau} & = \frac{1}{8 \pi G_3}\frac{\sqrt{\al^2 
+ \dot{R}^2}-\sqrt{1+\dot{R}^2}}{R} \label{S1} \\
S^{\theta}{}_{\theta} & = \frac{1}{8 \pi G_3}
\left(\frac{\ddot{R}}{\sqrt{\al^2 +\dot{R}^2}}-
\frac{\ddot{R}}{\sqrt{1+\dot{R}^2}}\right). \label{S2}
\end{align}

Assuming the shell is a perfect fluid, the specific physical form of
the linear boundary tensor is 
\begin{equation}
S^{a}{}_{b} = (\lambda +p) u^a u_b + p
h^{a}{}_{b}\,,
\label{perffluid}
\end{equation}
where $p$ is the linear pressure and $\lambda$ is the linear
mass density
of the shell.  So for a perfect fluid $S^{\tau}{}_{\tau} = -\lambda$
and $S^{\theta}{}_{\theta} = p$. Thus, from
Eqs.~(\ref{S1}) and (\ref{S2}), it follows that
\begin{align}
\lambda & = \frac{1}{8 \pi G_3}\frac{\sqrt{1
+ \dot{R}^2}-\sqrt{\al^2+\dot{R}^2}}{R} \\
p & = \frac{1}{8 \pi G_3}\left(\frac{\ddot{R}}{
\sqrt{\al^2 +\dot{R}^2}}-\frac{\ddot{R}}{\sqrt{1+\dot{R}^2}}\right)
\end{align} 
This set of equations has been previously found and studied 
in \cite{mannoh} in a 2+1 dynamical shell collapsing setting.
Taking the static limit $\dot{R} = \ddot{R} = 0$ in the above
equations and using the definition of the shell's rest mass 
\begin{equation}
M = 2 \pi
R \lambda\,,
\label{massdef}
\end{equation}
we have immediately
\begin{align}
M & = \frac{1-\al}{4 G_3} \label{m}\, \\
p & = 0. \label{m2}
\end{align}
To have a properly defined shell radius one has to impose
$\al>0$. Imposing positive mass it 
follows $0<\al<1$, whereas negative masses 
appear in the range 
$1<\al<\infty$.
From Eq.~(\ref{m2}) we see that in order for the shell to be static in a 
(2+1)-dimensional spacetime, its linear pressure must vanish.

We now want to study this static thin shell spacetime 
\cite{mannoh}
from a thermodynamics point of view. Through the formalism presented 
in \cite{york1} and developed by Martinez \cite{Mart}
for the thermodynamics of thin shells, we study the thermodynamics 
and find the entropy of
these thin shells in (2+1) flat dimensions.

\section{Thermodynamics, entropy equation for the shell, 
and stability}
\label{thermo}

We assume that the entropy of the shell can be expressed
in terms of its characteristics, namely, 
its rest mass $M$ and radius $R$. It is more useful, though
equivalent, to work with $M$ and $A$ \cite{Mart}, where $A$ is the 
area (here circumference) of the shell,  $A \equiv 2 \pi R$. 
Thus we write, 
\begin{equation}
S = S(M,A)\,.
\end{equation}
Assuming further 
that the shell is at some local temperature 
$T$, an explicit
expression for $S$
can then be found by directly integrating the first law of
thermodynamics
\begin{equation}
TdS = dM + p\, dA\,.
\end{equation}
By putting $p=0$ as given in Eq.~(\ref{m2}),
the first law for the (2+1)-dimensional thin shell 
spacetime simplifies to
\begin{equation}\label{1L}
T dS = dM\,.
\end{equation}
Equation (\ref{1L}) can be integrated
provided that the integrability conditions are satisfied.
In this case there is only one condition. It 
states that $T=T(M)$, i.e., the temperature $T$ is 
a function of the mass $M$ alone. 
This simple dependence stems from the 
fact that $p=0$ as given in Eq.~(\ref{m2}). 
Then, the
most general expression for $S$ is
\begin{equation}\label{SM}
S(M) = \int \be(M)dM + S_0
\end{equation}
where $\beta \equiv 1/T$ is the local inverse
temperature of the shell at the
equilibrium position $r = R$,
and $S_0$ is an integration constant.
Note that $\be(M)$ is an arbitrary
function of the mass that can be specified once
the specific matter
fields that constitute the shell are known.

Local intrinsic stability of the shell can also be studied. The former
is guaranteed as long as all the following inequalities are verified
\begin{equation}\label{C1}
\left(\frac{\dd^2 S}{\dd M^2}\right)_A \leq 0\,,
\end{equation}
\begin{equation}\label{C2}
\left(\frac{\dd^2 S}{\dd A^2}\right)_M \leq 0\,,
\end{equation}
\begin{equation}\label{C3}
\left(\frac{\dd^2 S}{\dd M^2}\right)\left(\frac{\dd^2 S}{\dd A^2}\right) 
- \left(\frac{\dd^2 S}{\dd M \dd A}\right)^2 \geq 0\,,
\end{equation} 
where the formalism developed in \cite{callen}
is being followed.

\section{Two specific equations of state for the 
thin shell matter: Entropy and 
stability}
\label{eqstate}

\subsection{The simplest equation of state}

The simplest form that can be considered for $\be(M)$ is a
power law,
\begin{equation}\label{f}
\be(M) = \g\, G_3^{(1+u)}M^u\,
\end{equation}
where we are assuming $M\geq0$, $\g$ and $u$ are free parameters
with $\g>0$ to guarantee positive temperature, the
factor $G_3$ must be present for dimensional reasons, and Boltzmann's
constant is set to $1$. In 3+1 dimensions, 
Planck's length $l_{\rm p}$,
with $l_{\rm p}=\sqrt{G_4\,h}$ ($G_4$
being the gravitational constant in 
four-dimensional spacetime and $h$ Planck's
constant),
appears
naturally in the temperature of a thin shell, since 
for a given mass there is always an intrinsic 
length associated with it (the gravitational radius of
the system), and so to have the correct units for 
the entropy one must resort to $l_{\rm p}$.
However, here in flat 2+1 dimensions there is no
intrinsic spacetime radius, and so Planck's length does not appear in
this analysis at all. This problem is thus purely classic and $G_3$,
with units of inverse mass, suffices to set the scale.

When $\be(M)$ has the form given 
in Eq.~(\ref{f}), one can
substitute this in (\ref{SM}) to get
\begin{equation}\label{S11}
S(M) = \frac{\g}{u+1}\left(G_3\,M\right)^{(1+u)} + S_0\,,
\,\quad {\rm for}  \; u \neq -1\,,
\end{equation}
and
\begin{equation}\label{S12}
S(M) = \g \ln(G_3\,M) + S_0\,,
\,\quad {\rm for}  \; u = -1\,.
\end{equation}

Although the values of the parameters $\gamma$ and $u$ cannot be
calculated without first specifying the nature of the matter fields,
it is possible to constrain them such that physical equilibrium states
of the shell are possible.  Starting with $S_0$, it is natural to
assume that a zero mass shell should have zero entropy, i.e., $ S(M
\to 0)=\int \be(M)dM + S_0 \to 0$.  It is seen directly from 
Eq.~(\ref{f})
that the entropy diverges when $M\to0$ for $u\leq-1$. Therefore the
above normalization condition can only be satisfied for $u > -1$ and
$S_0 = 0$.

In relation to stability, it is seen that conditions
(\ref{C2}) and (\ref{C3}) are automatically satisfied for any $u$. On
the other hand, one can find that condition (\ref{C1}) can
only be satisfied provided that $u\leq0$. Thus we conclude that
assuming a power law equation of the form (\ref{f}), stability of the
shell is possible for any $M\geq0$ as long as the parameter values of
$u$ are restricted to
\begin{equation}\label{u}
-1 < u \leq 0\,.
\end{equation}

One can also consider negative values of $M$, i.e., $\alpha>1$. The
relation (\ref{f}) would be of the same form with the proviso that one
takes the absolute values of $M$. The same results would follow.

\subsection{A more contrived equation of state}

Another possibility for $\be(M)$ could be a quadratic function in $M$,
of the form
$ \be(M) =
\de\, G_3^{(1+a)}(M+C M^2)^a
$,
where $\de$ and $a$ are some parameters with $\de>0$ to 
guarantee positive temperature, and $C$ is some constant. The
constant $C$, however, has a natural connection to 
Eq.~(\ref{m}). Indeed,
defining $m \equiv M + C M^2$ we can solve for $M$, obtaining the
physical solution
$ 
M = \frac{-1\\+\sqrt{1+4 C m}}{2 C}.
$ 
Comparing with Eq.~(\ref{m}), we can make the association $C = -2 G_3$ and
so $\al = \sqrt{1+4Cm} = \sqrt{1-8 G_3 m}$, thus arriving at the
natural quadratic expression for $\be$
\begin{equation}\label{b}
\be(M) = \de\,G_3^{(1+a)}(M - 2 G_3 M^2)^a.
\end{equation}
Equation (\ref{b}) is most easily integrated in the variable
$m$. Changing from $M$ to $m$ in Eq.~(\ref{SM}), defining the
parameter $\eta = \de/\al$, and changing back 
again to $M$, we obtain the entropy
\begin{align}\label{S21}
S(M) = \frac{\eta}{a+1}\, G_3^{(1+a)} (M - 2 G_3 & M^2)^{(1+a)} + S_0
\,,\nonumber\\
\,& \quad {\rm for}  \; a \neq -1
\end{align}
and
\begin{equation}\label{S22}
S(M) = \eta \ln \left[G_3\left(M - 2 G_3\, M^2\right)\right] + S_0
\,,
\,\quad {\rm for}  \; a = -1\,.
\end{equation}

Again, although the values of the parameters $\eta$ and $a$ cannot be
calculated without first specifying the nature of the matter fields,
it is possible to constrain them such that physical equilibrium states
of the shell are possible.  Starting with $S_0$, it is natural to
assume that a zero mass shell should have zero entropy, i.e., $ S(M
\to 0)=\int \be(M)dM + S_0 \to 0$.  It is seen directly from (\ref{b})
that the entropy diverges when $M\to0$ for $a > -1$. Therefore the
above normalization condition can only be satisfied for $a > -1$ and
$S_0 = 0$.

From Eq.~(\ref{b}), the stability equations
(\ref{C2}) and (\ref{C3}) are automatically satisfied since the
entropy does not depend on $A$.  Also, as mentioned above, $\al>0$ to
have a physical acceptable solution.  Considering Eq.~(\ref{C1}), it is
possible to show that it implies the inequality
\begin{equation}
(2a+1)\al^2 -1 \leq 0\,.
\label{aaa}
\end{equation}
We can study the two cases, $M>0$ and $M<0$.

For $M>0$, $\al$ is in the 
range $0 < \al \leq 1$. In this case Eq.~(\ref{aaa})
is automatically
satisfied if the exponent $a$ obeys
$a \leq 0$. It is also satisfied for $a
> 0$ but only if
$0<\al \leq \sqrt{\frac{1}{2a+1}}$.
Equivalently, this means that the rest mass $M$ of the shell must be
within the range 
\begin{equation}
\frac{1-\sqrt{1-2a/(2a+1)}}{4 G_3} < M \leq \frac{1}{4 G_3}\,,
\end{equation} 
for $a> 0$.

For $M<0$, we know that $\al>1$, and Eq.~(\ref{aaa}) requires 
additionally that $\al \leq \sqrt{\frac{1}{2a+1}}$. In terms
of rest mass $M$, this represents the range 
\begin{equation}
\frac{1-\sqrt{1-2a/(2a+1)}}{4 G_3} < M < 0.
\end{equation} 
However, since  $\sqrt{\frac{1}{2a+1}} > 1$ and $2a +1 >0$, we 
see that the analysis for $M<0$ is only 
valid for parameter values $-1/2 < a < 0$.

\section{Conclusions} 
\label{conc}

General relativity in a (2+1)-dimensional spacetime has no curvature
in empty space but in matter distributions curvature may still
exist. Einstein's equation thus still plays a role in determining the
required pressure and energy of a static thin shell (or ring in this
(2+1)-dimensional setting). 
Indeed, we have seen that in this situation the pressure must
be zero and the rest mass of the shell must satisfy Eq.~(\ref{m}).

Upon using the first law of thermodynamics we have found a specific
differential equation for the entropy of the ring that contained a
degree of freedom encoded in the inverse temperature $\beta$. We have
chosen the two simple ansatz for the inverse temperature, a power law
on the shell's rest mass and a quadratic form of it, obtaining two
distinct expressions for the ring's entropy.  This shell's entropy is
purely classic, as the only fundamental constant that enters into the
problem is the (2+1)-dimensional gravitational constant $G_3$, which
has units of inverse mass. This entropy could perhaps be explained 
quantically if the shell's mass is given in terms of the 
mass of its elementary constituents and a fundamental
theory for the mass of those constituents particles is at hand.
A thermodynamic stability analysis yielded
the range for the allowed parameters, revealing that the shell's rest
mass must be confined to a given interval if the shell is to be
stable.

Our results are of importance if one uses a concrete model for the
shell, e.g., a model 
involving fundamental scalar fields. One would then
extract from the model the specific equation of state with its precise
values for the exponents and constants, and could immediately
determine through the above equations whether it would be
thermodynamic stable.

We also note that these ring shell spacetimes when extended into 3+1
dimensions, through the use of a trivial coordinate $z$ say, represent
infinite cylinders. Thus this thermodynamic study also holds in 3+1
general relativity for those cylindrical thin shells.

\begin{acknowledgments}
We thank FCT-Portugal for financial support 
through Projects No.~PTDC/FIS/098962/2008 and No.~PEst-OE/FIS/UI0099/2011. 
\end{acknowledgments}

\end{document}